\documentclass[twocolumns,reprint,letterpaper,amssymb,superscriptaddress,nobibnotes, aps, prd]{revtex4-1}
\usepackage[dvipdfmx]{graphicx}
\usepackage{graphicx}
\usepackage{bm}
\usepackage{amssymb}
\begin{document}

\title{Measurement of the Newtonian Constant of Gravitation $G$ by Precision Displacement Sensors}

\author{Akio Kawasaki}
\email{akiok@stanford.edu}
\affiliation{W. W. Hansen Experimental Physics Laboratory and Department of Physics, Stanford University, Stanford, California 94305, USA}

\begin{abstract}
The Newtonian constant of gravitation $G$ historically has the largest relative uncertainty over all other fundamental constants with some discrepancies in values between different measurements. We propose a new scheme to measure $G$ by detecting the position of a test mass in a precision displacement sensor induced by a force modulation from periodically rotating source masses. To seek different kinds of experimental setups, laser interferometers for the gravitational wave detection and optically-levitated microspheres are analyzed. The high sensitivity  of the gravitational wave detectors to the displacement is advantageous to have a high signal-to-noise ratio of $10^{-6}$ with a few hours of the measurement time, whereas the tunability of parameters in optically-levitated microspheres can enable competitive measurements with a smaller scale setup dedicated to the $G$ measurement. To achieve an accuracy of $G$ better than currently available measurements, developments in force calibration is essential. These measurements can provide an alternative method to measure $G$ precisely, potentially leading to the improvement in the accuracy of $G$, as well as a better search for non-Newtonian gravity at a length scale of $\sim1$ m. 

\end{abstract}

\maketitle

\section{Introduction}
Measurements of fundamental constants of physics have been one of the most important work in the field of metrology. Over time, more and more precise measurements have been achieved \cite{CODATA2014}, affecting the definition of the units \cite{PhilTransRoyalSocA.363.1834}. Typically, these constants are measured with relative precisions around $10^{-7}$ or better. The Newtonian constant of gravitation $G$ has the worst relative precision of $4.7\times 10^{-5}$ \cite{CODATA2014}, leaving room for improvements. One of the reasons for the large uncertainty is discrepancies by a few standard deviations between different measurements. The classical device to measure $G$ is the torsion balance \cite{PhysRevLett.48.121,MeasTech.39.979,PhysRevLett.78.3047,PhysRevLett.85.2869,PhysRevLett.87.111101,PhysRevLett.91.201101,PhysRevD.71.127505,PhysRevD.82.022001,PhysRevLett.111.101102,PhilTransRoyalSoc.372.20140025,Nature.560.582}, as Cavendish used in 1798 \cite{Cavendish}. Other methods such as measuring weight change \cite{PhysRevD.74.082001}, atom interferometry \cite{Science.315.74,PhysRevLett.100.050801,Nature.510.518}, and displacement measurements with an optical cavity \cite{MeasSciTech.10.492,PhysRevLett.105.110801,Kleinevoss} have also been performed. All of these measurements so far have had uncertainties of at least 10 ppm \cite{Nature.560.582,CODATA2014}, and discrepancies in the value remain not only between different methods but also between the same methods performed by different collaborations. 

To improve the accuracy of the $G$ measurement, removing the discrepancies between different measurements is essential. The discrepancies are presumably caused by systematic errors that are not well estimated, which can have different sources in different experiments. One method to remove the discrepancies is to have a completely different setup that is potentially immune to the systematic errors previous measurements might have overlooked, as has long been discussed to increase the confidence level of overall measurements \cite{Nature.408.919}. In this paper, we propose a new way of measuring $G$ using precision displacement sensors. The method is to measure a periodic motion of a test mass that is driven by rotating source masses, and specifically, we analyze gravitational wave detectors and optically-levitated microspheres. The system is conceptually simple to analyze systematic errors, and has a potential for a $10^{-6}$ level signal-to-noise ratio (S/N). With a force calibration of similar precision, which requires developments in the future, a measurement of $G$  competitive to currently best measurements is possible. Because the detector and the source masses are completely separate, eliminating systematic background would be easier. The setup is also compatible with the test of non-Newtonian gravity, which is independent of the accuracy of $G$.

\section{Experimental Setup}
\begin{figure}[!tb]
	\begin{center}
 \includegraphics[width=0.6\columnwidth]{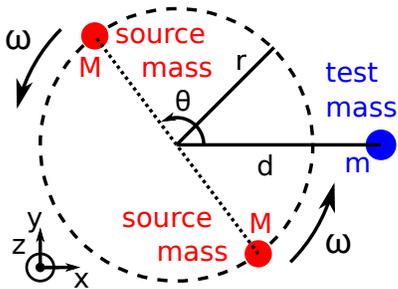}
 \caption{Experimental setup: a pair of source masses of mass $M$ are rotating at an angular velocity of $\omega$ on a circle of radius $r$ on each end of a line of $2r$ length. A test mass of mass $m$ whose position is monitored by a displacement sensor is located at $d$ away from the center of the circle.} 
 \label{ExpSetup}
 \end{center}
\end{figure}

Figure \ref{ExpSetup} shows the setup. It consists of a test mass and two source masses, all of which are assumed to be point masses for simplicity. The test mass of mass $m$ is an object whose displacement is monitored by a displacement sensor that generates a harmonic trap with a trapping frequency of $\omega_0$ and a damping constant of $\gamma$ for the test mass. Source masses of mass $M$ are on two opposite ends of a diameter of a circle on which the source masses rotate at an angular velocity of $\omega=2\pi f$. The distance between the test mass and the center of the circle is $d$, and for simplicity, the test mass is assumed to be located in the same plane as the circle. 

A force ${\bf F}=(F_{\rm x}, F_{\rm y})$ on the test mass due to the gravitational attraction by the source masses is calculated from the law of universal gravitation, with $k$ defined as $k=d/r$.
\begin{eqnarray}
F_{\rm x}&=&\frac{GMm}{r^2}\Big[ \frac{k-\cos \omega t}{(1+k^2-2k \cos \omega t)^{3/2}} \nonumber \\
& & + \frac{k+\cos \omega t}{(1+k^2+2k \cos \omega t)^{3/2}} \Big] \label{EqFx}\\
F_{\rm y}&=&\frac{GMm}{r^2}\sin \omega t \Big[ \frac{1}{(1+k^2-2k \cos \omega t)^{3/2}} \nonumber \\
& & + \frac{1}{(1+k^2+2k \cos \omega t)^{3/2}} \Big] \label{EqFy}
\end{eqnarray}
A calculation for an almost identical setup is performed in Ref. \cite{ClassQuantGrav.35.235009}, and their result corresponds to Eq. \ref{EqFx} in the limit of $k\gg1$. The time dependent position of the test mass is calculated by numerically integrating the equation of motion by the fourth order Runge-Kutta method. The initial position is set as the average position of the test mass determined approximately by the average force on the test mass by the source masses to keep a damped oscillation in the harmonic trap from being too large compared to the amplitude of the vibration induced by the source masses, together with the initial velocity of zero. In case a nonnegligible amount of damped oscillation remains, which happens only when $d$ and $r$ are extremely large in Fig. \ref{ScanDManyHarmonics}, the initial position is adjusted manually to remove it. The adjustment is at most $10^{-13}$ m. 

The fast Fourier transformation of the position of the test mass has higher order harmonics, because the force onto the test mass is not sinusoidal. Even with a reasonable estimate of the initial position, a small amount of relaxation of the test mass position to the equilibrium  point is unavoidable, which appears in the frequency domain as a smooth response function of a driven damped harmonic oscillator. To remove the contribution of this damped oscillation from the signal, the geometrical mean of the two adjacent bins of the signal bin is subtracted from the value of the signal bin. The Fourier transformed signal is compared with the sensitivity of the displacement sensors. Although motions in both the $x$ direction and the $y$ direction can be used for the measurement, only the $x$ axis is used in the following analysis. This does not loose generality, since $k\geq1$ and therefore $F_{\rm x}\gtrsim F_{\rm y}$. 

\section{Analysis on signal-to-noise ratio}
\subsection{Gravitational wave detetors}
First, gravitational wave detectors are analyzed. As a representative of several detectors \cite{PhysRevD.93.112004,ClassQuantGrav.32.024001,ClassQuantGrav.29.124007}, Advanced LIGO \cite{PhysRevD.93.112004} is used. To reduce unwanted perturbations to the detector system, it is assumed that the source masses are located at the far end of an arm. The test mass is a cavity mirror of $m=40$ kg \cite{PhysRevD.93.112004}. The $x$ axis is set along the cavity axis, and the source masses are on the far side of the cavity system. The motion of the mirror in the $x$ direction is governed by the horizontal pendulum mode of $\omega_0=2\pi \times 0.65$ Hz \cite{ClassQuantumGrav.29.235004}. To be conservative, the damping constant $\gamma$ is set as $2\pi \times 0.065$ Hz, which is derived from the lowered quality factor of $Q\sim10$ \cite{ClassQuantumGrav.29.235004}. Note that even if the actual quality factor is 1000 times larger than this value, the increase in the signal is only by a factor of 2.25, which happens at a 5 Hz rotation, and in a typical rotation frequency of 20 Hz that is used in the following analysis, the change in the signal is only 13 \%. Therefore, the change in $Q$ does not affect the following discussion more than a factor of few. Remaining free parameters of the system are $M$, $\omega$, $r$, and $d$. Here, $M=100$ kg is assumed. 

\begin{figure}[!tb]
	\begin{center}
 \includegraphics[width=1\columnwidth]{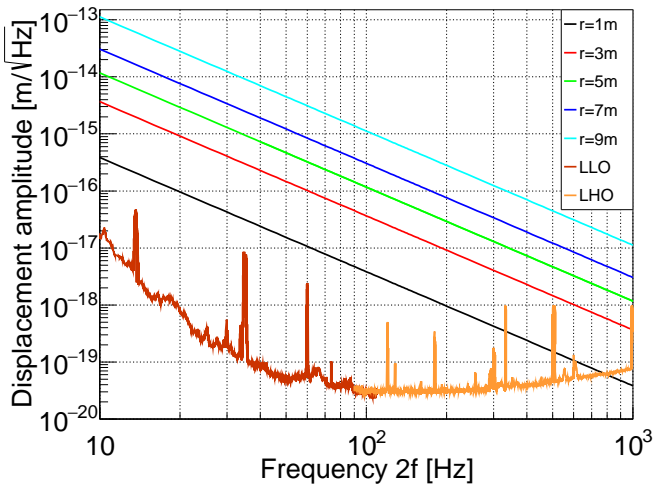}
 \caption{The amount of the signal at a frequency of $2f$ for different rotation frequencies $f$ of the source masses: straight lines from bottom to top show the signal at $r=1,~3,~5,~7,~{\rm and}~9$ m. $d$ is fixed at 10 m. The dark and light orange curves at the bottom are the sensitivity curves of LIGO Livingston Observatory (LLO) and LIGO Hanford Observatory (LHO), respectively, cited from Fig. 5 of Ref. \cite{PhysRevD.93.112004}.} 
 \label{ScanFManyR}
 \end{center}
\end{figure}

\begin{figure}[!tb]
	\begin{center}
 \includegraphics[width=1\columnwidth]{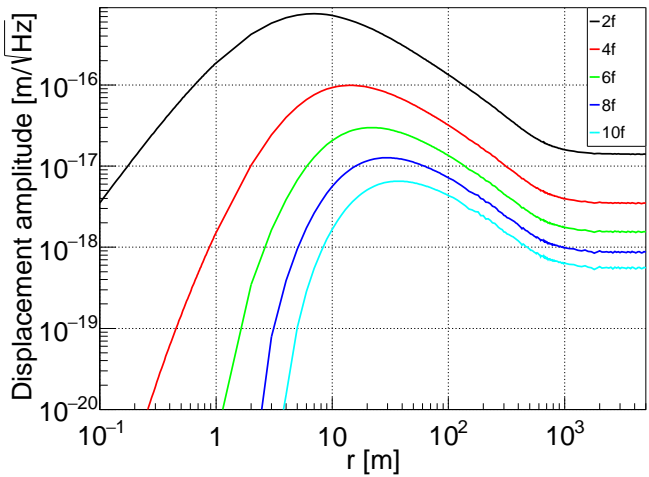}
 \caption{The behavior or higher harmonics when $r$ is changed: lines from top to bottom shows the signal at the frequency of $2f$, $4f$, $6f$, $8f$, and $10f$. $d-r$ and $f$ are fixed at 5 m and 20 Hz, respectively. Note that $r\gtrsim 10$ m is not realistic to construct, and the large radius region is shown here only for the purpose of displaying an asymptotic behavior. } 
 \label{ScanDManyHarmonics}
 \end{center}
\end{figure}

Because there are two source masses, the basic behavior of the test mass is an oscillation at a frequency of $2f$. When $f$ increases, the signal decreases proportional to $f^{-2}$ (see Fig. \ref{ScanFManyR}). This behavior is true for all harmonics, and higher harmonics have a smaller signal, as Fig. \ref{ScanDManyHarmonics} shows. The comparison between the signal and the sensitivity curve of Advanced LIGO \cite{PhysRevD.93.112004} in Fig. \ref{ScanFManyR} tells that the signal around the frequency $2f\simeq40$ Hz has the best S/N. Also, at $f\simeq20$ Hz, a few higher harmonic signals are also above the noise level of Advanced LIGO. When $r=5$ m and $d=10$ m, the S/N is more than 1 for up to 6th harmonics. 

When $d$ is fixed, larger $r$ gives a larger signal (Fig. \ref{ScanFManyR}), because the difference between the minimum and the maximum distance between a source mass and the test mass ($d-r$ and $\sqrt{d^2+r^2}$, respectively) is larger when $d-r$ is smaller. The minimum distance $d-r$ is practically determined by the size of the objects around the test mass, such as the test mass and the source masses themselves with finite sizes, the damping system, and the vacuum chamber. A rough estimate of the size of all objects surrounding the test mass is a couple of meters, based on dimensions shown in Ref. \cite{ISMA2014}, and therefore having a minimum distance of 5 m between the test mass and the source masses, which is $d-r$ , would accommodate all objects originally located around the test mass. In the following discussion, $d-r=5$ m is assumed, but smaller $d-r$ increases the signal significantly. 

When $d-r$ is fixed, remaining parameter is $r$. Figure \ref{ScanDManyHarmonics} shows the amount of signal for different $r$ with $d-r=5$ m and $f=20$ Hz. When $r$ is small, the difference between the maximum force $F_{\rm max}$ and the minimum force $F_{\rm min}$ on the test mass is small, and therefore the signal is small. In the limit of $d\gg r$, the force is approximately 
\begin{equation}
F_{\rm x}=2\frac{GMm}{d^2}\left(1-\frac{3}{4}\frac{r^2}{d^2}+\frac{9}{4}\frac{r^2}{d^2}\cos 2 \omega t \right),
\end{equation}
which is purely sinusoidal. This explains small higher harmonic signals at small $r$. At the limit of $r\rightarrow \infty$, $F_{\rm max}-F_{\rm min}$ is large, but the velocity of a source mass to pass the closest point to the test mass is high, and overall work on the test mass by the source masses becomes small. Also, most of the force applied on the test mass occurs when the source mass is closest to the test mass, i.e. $\theta \ll 1$. Since $\cos \theta=\cos \omega t$,
\begin{equation}
F\simeq \frac{GMm}{d^2}\frac{\Delta + (\omega t)^2}{(\Delta^2 + 2 \Delta (\omega t)^2 + (\omega t)^2)^{3/2}},
\end{equation}
when $\omega t\ll 1$, where $\Delta=d-r$. This is independent of $r$, but still the force is a function of $\omega t$, and therefore higher harmonic signals converge to constants. Thus, there is an intermediate $r$ that maximizes the signal, which is around $r=8$ m. On one hand, larger higher harmonics help determining $G$ more precisely, because we can use more than one signal frequency. On the other hand, higher harmonic signals are at least a factor of few smaller than that of the basic harmonics, and therefore optimizing $r$ for obtaining a large signal for higher harmonics might not necessarily benefit a lot. 

The basic harmonic signal is significantly larger than the noise level of LIGO. Take $r=5$ m, $d=10$ m, $f=20$ Hz, and $M=100$ kg Hz as an example. The signal is $7.3\times10^{-16}$ m/$\sqrt{\rm Hz}$, which results in  ${\rm S/N}=7.3\times10^3$. With this amount of signal, only $\sim200$ s integration time reaches the S/N same as the precision of currently most precise measurements, and in a few hours, the S/N increases to $10^{-6}$. It should be noted that to measure the value of $G$ down to a precision of $10^{-6}$, all of the quantities in the equation $F=GMm/r^2$ have to be measured with a precision of $10^{-6}$. This means the force also has to be measured with this high precision. Methods to improve the precision of the force calibration is discussed in Section \ref{calibration}. 

\begin{figure}[!tb]
	\begin{center}
 \includegraphics[width=1\columnwidth]{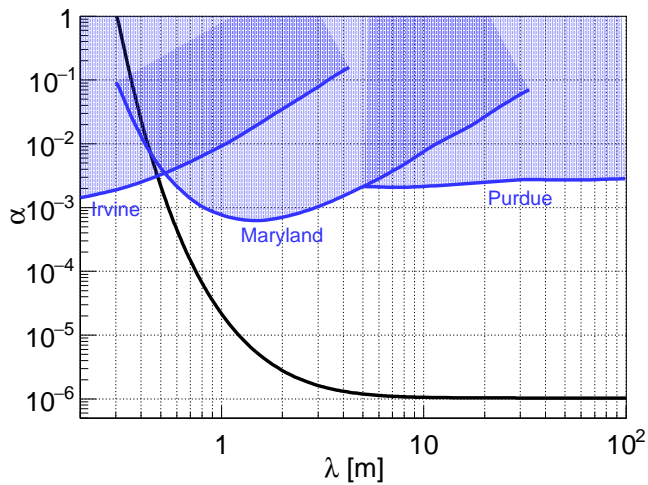}
 \caption{A potential constraint on non-Newtonian gravity by the system described in Fig. \ref{ExpSetup} with a $10^{-6}$ precision force measurement (black curve): non-Newtonian gravity is assumed to have a form of the Yukawa potential, and the plot is made in the two dimensional space of relative magnitude $\alpha$ of the non-Newtonian term compared to the Newtonian term and the length scale $\lambda$ of the exponential decay of the Yukawa potential. Blue curves with shaded region are current limit adopted from Ref. \cite{ClassQuantumGrav.32.033001}. Original data comes from Ref. \cite{PhysRevD.32.3084} (Irvine), Ref. \cite{PhysRevLett.70.1195} (Maryland), and Ref. \cite{PhysRevLett.61.1159} (Purdue). } 
 \label{NNG}
 \end{center}
\end{figure}

If the measured displacement agrees with the theoretical estimate based on Eqs. \ref{EqFx} and \ref{EqFy} up to a $10^{-6}$ precision, the assumption of the inverse square law is correct to the same precision. This serves as a search for non-Newtonian gravity. Assuming the following modification by a Yukawa potential term with dimensionless magnitude $\alpha$ and length scale $\lambda$, 
\begin{equation}
V(r)=\frac{GMm}{r}\left(1+\alpha e^{-r/\lambda}\right)
\end{equation}
the expected sensitivity in $\alpha-\lambda$ space is depicted in Fig. \ref{NNG}, which is beyond the current best constraint at $\lambda\gtrsim1$ m. Note that the search for the non-Newtonian gravity is independent of the accurate measurement of $G$, whose difficulties are discussed in Section \ref{Accuracy}, because accurate extraction of the contribution of $M$ and $m$ on $GMm$ is not necessary to extract the effect of the Yukawa potential term.

The argument so far suggests that $8\lesssim r \lesssim 20$ m is optimal. The practical limit for $r$ is determined by the power of the motor that moves the turn table where the source masses are mounted. At the parameters of $d=10$ m, $r=5$ m, $f=20$ Hz, and $M=100$ kg, the total kinetic energy of the source masses is 40 MJ. To attain this, a relatively high power motor or engine is necessary. (for example, an airplane is $\sim1$ GJ and a car is $\lesssim 1$ MJ) The centripetal force required to support a source mass with these parameters is $7.9 \times 10^6$ N. Assuming the two source masses are connected by a beam of 304 stainless steel, whose ultimate tensile strength is 520 MPa, a beam of 7 cm radius barely supports a source mass. The drag force by the air is $1.2 \times 10^4$ N, with an assumption that the source mass is made of a lead sphere, resulting in the power of 7.8 MW to maintain the angular velocity. This is an order of magnitude larger than a high-power car engine, and therefore it is essential to move the source masses in a vacuum or making a wind shield to reduce the drag force. Potentially, it is more realistic to reduce the $r$ or $f$. $r=1$ m requires a smaller kinetic energy, centripetal force and power by a factor of 25, with a scarification of the signal by a factor of 3. Smaller radius for the source masses' rotation is also beneficial if the space around the test mass mirror is limited. In case there is no room to put an additional device around the mirror, the space can be reserved in the next generation gravitational detector \cite{ClassicalQuantumGrav.27.194002}. Because of an order of magnitude improvement in the position sensitivity in the next generation detector, even shorter integration time is expected. 

\subsection{Optically-levitated microspheres}
\begin{figure}[!tb]
	\begin{center}
 \includegraphics[width=1\columnwidth]{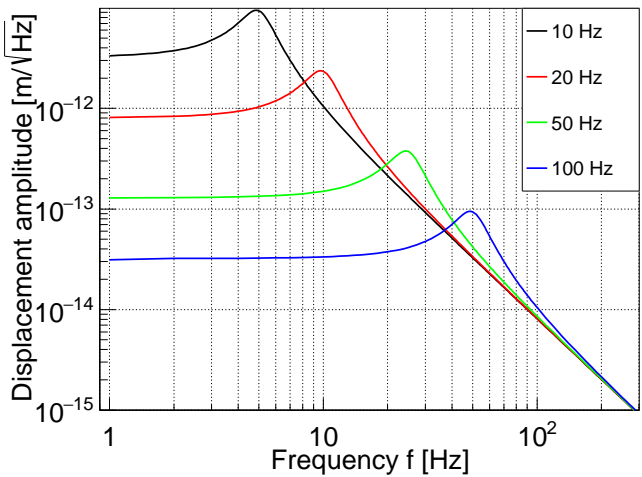}
 \caption{The amount of the signal at the frequency of $2f$ for different rotation frequencies $f$ of the source masses: lines from bottom to top show the case of $\omega_0/2\pi=10,~20,~50,~{\rm and}~100$ Hz. $d$ and $r$ are fixed at 0.6 m and 0.3 m, respectively. $Q$ is fixed at 3. } 
 \label{BeadScanFManyF0}
 \end{center}
\end{figure}

The optically-levitated microsphere is another system suitable for precision displacement sensors. Although they have sensitivities to the displacement around $10^{-11}$ m/$\sqrt{\rm Hz}$ \cite{PhysRevA.97.013842,PhysRevA.96.063841,NatPhys.7.527,PhysRevA.93.053801}, poorer than the gravitational wave detectors, they have advantages of compact sizes and tunability of parameters for the harmonic traps, $\omega_0$ and $\gamma$. The resonant frequency $\omega_0/2\pi$ varies from 20 Hz \cite{ApplPhysLett.28.333} to a few kHz\cite{PhysRevA.93.053801}. $\gamma$ is typically set at the same order of magnitude as $\omega_0$ to keep the residual noise small and to generate large enough damping to keep the sphere in the trap at high vacuum, but alternating cycles of strong and weak feedback cooling can achieve a situation of a low noise and small $\gamma$. The dependence of the signal on $f$  for different $\omega_0$ is shown in Fig. \ref{BeadScanFManyF0}. The signal is a standard harmonic oscillator driven at a frequency of $2f$.

The amount of the signal at $f=\omega_0/2\pi$ increases linearly with $Q$. The optimal value for $Q$ is determined by a compromise between the amount of the signal, the amount of the noise, and the time required for cooling and equilibration. At $f=\omega_0/2\pi$, not only the signal but also the noise increases, compared to off-resonant frequencies, if the dominant noise is induced by a force of white noise, e.g. background gas collision. To remove the noise, the measurement sequence can include a periodic cooling stage with lower $Q$, during which the response of the test mass is reduced due to the larger damping constant $\gamma$ but the noise is reduced to the noise floor. At high $Q$, the time necessary for the driven oscillation to reach the large equilibrium amplitude can be too long to keep the noise level low. This puts an upper limit on $Q$. These factors should give an optimal $Q$, which is at least 3, as the typical operation of Ref. \cite{PhysRevA.97.013842} is performed around $Q=3$. 

Other than the resonant frequency, the behavior of the optically-levitated microsphere is the same as that of the gravitational wave detectors. Above the resonant frequency, lower $f$ leads to a higher signal, meaning that a system with a lower resonant frequency gives a higher signal. This motivates construction of a low trapping frequency system. Even at $\omega_0=2\pi \times 10$ Hz, the position sensitivity of the currently available system \cite{PhysRevA.97.013842} has at best S/N of 1. Further reduction of the noise is desired for better S/N, and if it decreases to $10^{-13}$ m/$\sqrt{\rm Hz}$, which is the shot noise limit of radial directions in Ref. \cite{PhysRevA.97.013842}, S/N becomes 100. With this, an integration time of $10^6$ s, which is a couple of weeks, brings the similar S/N to the currently available best precision. 

Other geometrical configurations can improve the measurement. Although a single source mass would not be a good choice due to the large vibration at frequency $f$ generated by the mass imbalance, four or larger even number source masses can be beneficial. On one hand, this reduces the signal, because $F_{\rm max}-F_{\rm min}$ decreases, but on the other hand, reduced $f$ to obtain a signal at the same frequency is advantageous to reduce the energy required to drive the rotation of the source masses. Because the kinetic energy scales $\sim f^2$, increasing the number of source masses by a factor of $n$ reduces the total kinetic energy to $1/n$. An advantage specific to optically-levitated microspheres is the force measurement is performed three-dimensionally \cite{PhysRevA.99.023816}. This allows the use of $F_{\rm y}$, which potentially helps increasing the precision of the measurement and discriminating background. If the trapped sphere is off the plane of the source masses' rotation, $F_{\rm z}$ is also non-zero, and this also provides a signal. 

\section{Analysis of background}
The discussion so far only described the S/N and the background has not been considered yet. Because the signal is a continuous oscillation at a frequency $2f$, all of transient backgrounds can be removed by taking long enough measurements and picking up a specific bin in the frequency space. Monitoring environmental vibrations and vetoing the data acquisition when the environmental vibration is large also works. If necessary, it is even possible to put rotating source masses at more than one gravitational wave detector sites to take coincidence of the two independent detectors to reject phase-incoherent signals as backgrounds, in the same way as the gravitational wave detection. 

The most difficult background to remove is the vibration induced by the motion of the source masses. The structure is ideally symmetric relative to the center of the circular orbit, but a finite mechanical tolerance can induce the shift of the center of mass of two source masses from the center of the circle. This generates mechanical oscillations at frequency $f$, and the nonlinear response of the holding structure easily generates vibrations at $2f$. To suppress these vibrations, precise tuning of the center of mass needs to be performed. This can be achieved by adjusting the position of screws in screw holes. As there can be multiple screws on different positions in the rotating structure, both coarse and precise tuning of the center of mass is possible. For example, when $M=100$ kg, a screw of 10 g can tune the center of mass by 1 $\mu$m, with a motion of 1 cm. Presumably, this screw feature should be on a holding structure, not in the source masses, as source masses themselves have to be uniform to accurately calculate the gravitational force when the finite size is incorporated to the calculation.

Remaining vibrations after carefully tuning the center of mass should be isolated by a vibration isolation system. The ultimate goal of the isolation is to reduce it below the ground noise, which is $\sim 10^{-9}$ m/$\sqrt{\rm Hz}$ at 20 Hz \cite{PrecisionEngineering.40.287}. However, because the seismic vibration at 20 Hz is $\sim30$ times below the sensitivity of the detector, the realistic goal of the vibration isolation is to reduce the vibration by the source masses transmitted to the ground down to the $3 \times 10^{-8}$ m/$\sqrt{\rm Hz}$. Vibration isolation systems for $>100$ kg objects are widely available. For example, Advanced LIGO's in-vacuum seismic isolator (ISI) \cite{PrecisionEngineering.40.273} has transfer function of $1\times 10^{-7}$ m/N, with a capacity of 1000 kg. If we suppose the remaining offset of the center of mass is 1 $\mu$m, source masses rotating at 20 Hz generates $\sim3$ N. Something similar to the ISI results in a vibration of $\sim 3 \times 10^{-7}$ m amplitude, which is still an order of magnitude larger than the targeted vibration level. Stacking another layer of vibration isolation system reduces the amount of the vibration by one or two orders of magnitude. An additional vibration isolation stage with a capacity of $\sim1000$ kg can be chosen from commercially available products, or can be designed specifically for the system. A magnetic bearing for the axis of the rotating source masses might help reducing the vibration, as well as reducing the friction.  It should be mentioned that the operation of the vibration isolation system discussed here is opposite from the standard vibration isolation for LIGO. In the case discussed here, the vibration at the top of the stage needs to be reduced at the bottom of the stage mounted onto the ground, whereas LIGO requires the vibration at the ground should be decoupled from the stationary mirror. Because the basic principle of vibration isolation is a spring with low resonant frequency, it should work in both ways. Also, note that the discussion here is the suppression of the vibration at $f=20$ Hz, and the vibration at the signal frequency $2f=40$ Hz should be smaller than that at 20 Hz. This also contributes to the suppression of the background at the signal frequency. 

The displacement of the source masses due to the vibration itself needs to be well monitored to detect the relative change of the position by 1 $\mu$m, which is $\lesssim10^{-6}$ of the distance between the source masses and the test mass. State-of-the-art interferometric position sensors easily perform position sensing well below this requirement \cite{SensActA.190.106}. Another uncertainty can come from geometrical imperfection. Precision machining typically has a precision of submicrons, so to the extent that $r$ and $d$ are in the order of 0.1 m or larger, the relative uncertainty can be suppressed to $\sim 10^{-6}$. The precision of mass measurement can also be as high as $10^6$. Overall, suppressing the background and geometrical systematic uncertainty down to $10^{-6}$ level seems possible. 

\section{Force calibration}\label{calibration}
Because the goal of the $G$ measurement is to perform the measurement of the displacement of the test mass when a theoretically known amount of force is applied, force calibration, i.e. examining the displacement of the test mass when experimentally known amount of force is applied, is essential. If the target precision of $G$ is $10^{-6}$, force calibration also has to have a precision of $10^{-6}$. Currently, LIGO collaboration performs it by changing the amount of radiation pressure (photon calibration), resulting in the precision of few percents. Even if there can be some technical developments in the photon calibration, improvement by four orders of magnitude would not happen soon. Newly proposed calibration method for the gravitational wave detectors uses the universal law of gravitation \cite{ClassQuantGrav.35.235009}, and therefore this method cannot be used for the $G$ measurement. 

If the system can be described by a combination of the harmonic oscillators orthogonal to each other, the difficulty of measuring the force with $10^{-6}$ precision can be avoided. Suppose a driving force is turned off, and the oscillator starts to behave as damped harmonic oscillator ($\omega>\gamma$). The motion of the oscillator is described as
\begin{equation}
x(t)=\left(A e^{i \sqrt{\omega_0^2-\gamma^2}t} +B e^{-i \sqrt{\omega_0^2-\gamma^2}t}\right)e^{-\gamma t}.
\end{equation}
$\omega_0$ and $\gamma$ are obtained from experimental data by fitting it with this equation. With these parameters known, response to the force $F(t)=mF_0 e^{-i\omega t}$ is 
\begin{equation}
x(t)=\frac{F_0}{(\omega^2-\omega_0^2)-i 2\gamma \omega} e^{-i\omega t}. \label{DrivenDampedOsci}
\end{equation}
In practice, there are more than one eigenmodes that couple to the signal direction. In this case, $\omega_0$ and $\gamma$ should be measured for all relevant eigenmodes.
As far as the system is in the region of linear response, simply multiplying Eq. \ref{DrivenDampedOsci} for different eigenmode with a coupling constant to the motion of our interest gives the overall response. 
The major difficulty in this method is that the resonant frequency of the most relevant eigenmode is 0.65 Hz, and reaching $10^{-6}$ precision for this low frequency requires $\sim10^6$ s measurement time. 

Another method, which is more suitable for optically-levitated spheres than gravitational wave detectors, is to apply the force through electric field. Because the force sensitivity of the system is $\sim10^{-17}$ N or less \cite{PhysRevA.93.053801,PhysRevA.99.023816}, the system is sensitive to the change of the amount of charge $q$ by the elementary charge $e$. This allows exact determination of the charge on the test mass, and all uncertainty of the force  $F=qE$ applied  by an external electric field $E$ comes from that of electric field. As there exists some DC power supply whose relative stability is in the order of 1 ppm, it can be possible to have force calibration with a precision below $10^{-5}$. Once the voltage is stabilized and determined remaining systematic error in the amount of electric field is the edge effect of the plates of electrode. The general strategy to reduce it is to have large enough plates, and locate them as close to each other as possible and as parallel as possible. Further estimate would require finite element analysis of the electric field, which is system specific and therefore beyond the scope of this paper. In case the response of the system is not linear beyond certain range of the applied force, we can stay in the linear range, and increase the S/N by integrating for a long time. For optically-levitated microsphere systems, it is relatively easy to implement this feature, as such electrodes to control force on the spheres is already implemented \cite{PhysRevA.99.023816,PhysRevLett.113.251801}. For gravitational wave detectors, there should be an additional feature to mechanically couple the mirror and the region where the charge is well controlled. 

\section{Discussion on accuracy}\label{Accuracy}
The most difficult part of the measurement of $G$ is not the precise measurement of the force but accurate removal of all systematic shifts. This means that the accurate measurement of $G$ down to the $10^{-6}$ level is not guaranteed, even if the S/N and the force calibration down to $10^{-6}$ is achieved. The advantage of the proposed systems is that the measurement method for the force is completely different from the conventional methods. The laser interferometry for gravitational wave detection and optically-levitated microspheres have never been used for the $G$ measurement, and therefore it can be immune to the systematic errors that are specific to other conventional force detectors. For example, torsion balances have systematic effects due to the suspension system. LIGO system does not involve torsional degree of freedom to its position detection, and optically-levitated microspheres are supported against gravity by a completely different method. 

Another advantageous feature of the proposed systems is that the test mass and source masses are completely separate. In the conventional methods, particularly for torsion balance experiments, force sensors are in the middle of turntables for the test masses, and they are close to each other. In Ref. \cite{Nature.560.582}, the distance between the center of the test mass and the source masses is $\sim 17$ cm \cite{RevSciInstr.85.014501}. However, in the case of gravitational wave detectors and optically-levitated microspheres, the typical smallest distances in the analysis above are 5 m and 30 cm, respectively. First, this means it is easier to put shielding objects between the test mass and the source masses. This reduces the electrostatic and magnetic effects. Second, the larger distance makes the system closer to the assumption of point mass in the calculation in Eqs. \ref{EqFx} and \ref{EqFy}. This reduces the systematic errors due to the imperfection in the source masses, such as deviations from the complete sphere, and the nonuniform density. This effect is significant for optically-levitated microspheres, which has a tiny test mass (e.g. 4.7 $\mu$m diameter in \cite{PhysRevA.99.023816}, 300 nm diamter in \cite{PhysRevA.93.053801}), where approximation to the point mass works well down to 10 ppm or less of the distance between the test mass and the source masses. Also, asymmetries in the shape can be averaged out by spinning the sphere along an $z$ axis in Fig. \ref{ExpSetup} \cite{PhysRevA.99.041802,PhysRevA.97.051802,PhysRevLett.121.033602}. In the case of gravitational wave detectors, the size of the test mass is $\sim 10$\% of the minimum distance, which is an order of magnitude smaller than Ref. \cite{Nature.560.582}. On one hand, this would eliminate the systematic error to some extent, but on the other hand, the fact that the mirrors of the gravitational wave detector is not optimized for the $G$ measurement can add some other systematic errors, and can require careful calculations that takes the mirror's actual shape into account for removing such effects. 

There are some disadvantages in the proposed systems, compared to the conventional methods. A major downside is that large size, which makes it difficult to precisely and uniformly stabilize the temperature of the whole system, which affects the distance between the test mass and the source masses due to the thermal expansion. Also, heavier objects induce a larger amount of deformation in the supporting system, which can generate additional uncertainty in the position. 

The choice of material for the source masses would depend on a compromise between the size and the uniformity. If a high density material such as lead is used, the size of the source mass becomes smaller, which is closer to the point mass approximation, but lead is not a material that has best uniformity or that is easy to machine. The best choice based on these two would be silicon, whose mass can be measured with 10 ppb precision and deviation from perfect sphere is 1 ppm level \cite{Metrologia.48.S1}. In the case of silicon, however, the radius for a 100 kg source mass is 21.7 cm, which is too large for bringing it as close to an optically-levitated microsphere as 30 cm. In this case, high density material with decent rigidity would be, for example, is tungsten. Its density 19.3 g/cm$^3$ results in 10.7 cm radius.
Also, the structure around the turntable has to be carefully designed to eliminate all possible systematic modulation in the signal due to its deviation from the symmetric structure. 

\section{Summary}
The precise detection of the displacement of a test mass can be a new method of measuring the Newtonian constant of the gravitation $G$. With the state-of-the-art gravitational wave detectors such as LIGO and a pair of 100 kg source masses rotating on a circular trajectory of radius 5 m, whose center is 10 m away from the test mass mirror, the S/N of $10^{-6}$ can be reached in a few hours of the integration time. The optically-levitated microspheres need certain amount of improvement in the sensitivity, and if the noise level decreases to $10^{-13}$ m/$\sqrt{\rm Hz}$, few weeks of the integration time can provide a similar S/N to the current best uncertainty. With a proper force calibration method, which has to be newly developed, these can contribute to a more accurate measurement of $G$, as the source of systematic shifts is different from conventional measurement methods. Also, even if the measurement of $G$ has some difficulties, proposed systems can be used for the search for non-Newtonian gravity in the length scale of $\sim 1$ m.

\section*{Acknowledgment}
The author acknowledges the partial support of a William M. and Jane D. Fairbank Postdoctoral Fellowship of Stanford University.

\bibliography{MeasuringG}

\end{document}